\documentstyle[12pt]{article}

\begin{document}

\title{\noindent {\small USC-97/HEP-B6 \hfill hep-th/9710066} \\
{\vskip 1.0cm} Gauge Principles \\
for Multi-superparticles{\thanks{{\protect\small Research partially
supported by the U.S. Department of Energy under grant number
DE-FG03-84ER40168, and by the National Research Council under grant number
GAC021197.}}} }
\author{{\bf I. Bars and C. Deliduman } \\
Department of Physics and Astronomy\\
University of Southern California\\
Los Angeles, CA 90089-0484}
\date{}
\maketitle

\begin{abstract}
\noindent We formulate new gauge principles for $n$ supersymmetric particles
in a worldline formalism with $N$ supersymmetries. The models provide
realizations of the more general supersymmetries that are encountered in
sectors of S-theory or Matrix theory, with a superalgebra of the form $%
\left\{ Q_\alpha ^A,Q_\beta ^B\right\} =\delta ^{AB}\,\gamma _{\alpha \beta
}^{\mu _1\cdots \mu _n}(p_{\mu _1}^1\cdots p_{\mu _n}^n).$ Due to the local
gauge and kappa symmetries the $n$ superparticle momenta $p^{i\mu }$ and $N$
supercharges $Q_\alpha ^A$ are constrained by $p^i\cdot p^j=0$ and $\gamma
\cdot p^iQ^A=0.$ The constraints have solutions only in a space with $n$
timelike dimensions and SO$\left( d+n-2,n\right) $ spacetime symmetry. The
cases SO(9,1), SO(10,2) and SO(10,3), with one, two and three timelike
dimensions respectively, are of special interest. In each case, due to the
constraints, the classical motion and quantum theory of each superparticle
are equivalent to the physics with a single time-like dimension in an
effective 10D superspace with SO$\left( 9,1\right) $ Lorentz symmetry.
\end{abstract}

\vfill\eject

\section{Multi-particles and multi-times}

Relativistic particles with spacetime symmetry SO$\left( d-1,1\right) $
which move freely in $d$ dimensions can be described by an action for $n$
worldlines $x_i^\mu \left( \tau \right) $ 
\begin{equation}
S=\frac 12\int_0^Td\tau \,\sum_{i=1}^n[e_i^{-1}\dot{x}_i^2-e_i\,m_i^2]\,\,.
\end{equation}
The einbeins $e_i\left( \tau \right) $ impose the mass-shell constraints $%
p_i^2+m_i^2=0$ on the canonical momenta $p_i^\mu =e_i^{-1}\dot{x}_i^\mu $
which are conserved according to the equations of motion $\partial _\tau
p_i^\mu =0.$ The constraints emerge because the action is gauge invariant
under the following $n$ independent infinitesimal reparametrizations 
\begin{equation}
\delta x_i^\mu \left( \tau \right) =\varepsilon _i\left( \tau \right)
\,\,\partial _\tau x_i^\mu \left( \tau \right) ,\quad \delta e_i\left( \tau
\right) =\partial _\tau \left( \varepsilon _i\,e_i\right) \left( \tau
\right) .  \label{gauge1}
\end{equation}

In recent papers \cite{twotimes}\cite{spartstr} a generalization of the
gauge symmetry (\ref{gauge1}) was introduced. The purpose of the present
paper is to propose an elegant and more satisfactory formulation of the
gauge symmetries and supersymmetries in \cite{twotimes}\cite{spartstr}\cite
{sezgin3}. Before we present the new formalism let us recall the role of the
gauge symmetry by reviewing the simplest two particle case. The new symmetry
requires orthogonal on shell momenta 
\begin{equation}
p_1^2+m_1^2=0,\quad p_2^2+m_2^2=0,\quad p_1\cdot p_2=0.
\end{equation}
This set of constraints have a solution provided there are at least two
timelike dimensions with SO$\left( d-1,2\right) $ spacetime symmetry if the
particles are massive, or SO$\left( d,2\right) $ if they are massless.
However, in the background of the other particle, each particle effectively
moves in a subspace with a single timelike dimension with SO$\left(
d-1,1\right) $ spacetime symmetry \cite{twotimes}.

The motivation for a higher geometrical space, with more than one timelike
dimension, comes from theoretical attempts to explain the supersymmetry and
duality symmetries among p-branes \cite{ibjapan}-\cite{14D}. As argued in 
\cite{twotimes}\cite{spartstr}, the presence of two or more time-like
dimensions is compatible with standard physics. The essential point is that
each particle moves effectively as if there is a single time-like dimension,
and satisfies the standard energy-momentum relation. When the particles (or
p-branes) of type 2,3,$\cdots $ are frozen to one energy-momentum eigenvalue
(i.e. Kaluza-Klein type reduction) the remaining theory describes the
physics for particles (or p-branes) of type 1. This leads to a cosmological
scenario for deriving a universe with an effective single timelike dimension
from a pre-Big-Bang universe with multi-timelike dimensions \cite{twotimes} 
\cite{spartstr}. In this scenario, the multi-times could be physically
relevant before the Big-Bang in an era of unified symmetries, including
supersymmetry and dualities of various p-D-branes \cite{14D}. After the Big
Bang phase transition, the symmetries are broken, and one species of
particles propagate in the background of all the others. Thus, the particles
that make up our present visible universe behave effectively as if there is
a single time-like dimension. Nevertheless, the compactified part of the
universe, with its multi-times, has an effect in determining the properties
and quantum numbers of low energy physics in new ways, as illustrated in 
\cite{superpv}.

\section{Multiparticle gauge symmetry}

The two particle constrained system can be generalized to an $n$ particle
constrained system of the form 
\begin{equation}
p^i\cdot p^j+ m_i^2 \delta^{ij}=0.  \label{multi-p}
\end{equation}
To have a solution with an effective SO$\left( d-1,1\right) $ symmetry for
each particle, one adds $\left( 0,1\right) $ dimensions for each additional
massive particle and $\left( 1,1\right) $ dimensions for each additional
massless one. Then there are $n$ timelike dimensions with spacetime symmetry
that ranges from SO$\left(d-1,n\right) $ if all particles are massive, to SO$%
\left( d+n-2,n\right) $ symmetry if all particles are massless. An action
that gives this set of constraints can be constructed by using $\tau $%
-independent {\it global} Lagrange multipliers, as in \cite{twotimes}\cite
{spartstr}. Here we will propose yet another approach using $\tau $%
-dependent {\it local} gauge functions, bringing the formulation closer to
standard concepts of local gauge and reparametrization symmetries. The
approach introduces new types of gauge symmetries for multi-particles as
well as new actions.

Consider the following action for $n$ particles described by $x_i^\mu \left(
\tau \right) $ 
\begin{equation}
S=\frac 12\int_0^Td\tau \,\left( \dot{x}_i^\mu e^{ij}\dot{x}_j^\nu \eta
_{\mu \nu }-\left( m^2\right)^{ij}e_{ij}\right),
\end{equation}
where $e_{ij}\left( \tau \right) $ is a symmetric matrix that generalizes
the einbein, and $e^{ij}\left( \tau \right) =\left( e^{-1}\right) ^{ij}$ is
its inverse. This action is invariant under the following gauge symmetry
that replaces the standard $\tau $-reparametrizations of (\ref{gauge1}) 
\begin{equation}
\delta x_i^\mu =\varepsilon _{ij}\left( \tau \right) e^{jk}\left( \tau
\right) \,\partial _\tau \,x_k^\mu \left( \tau \right) ,\quad \delta
e_{ij}=\partial _\tau \varepsilon _{ij}\left( \tau \right) .  \label{gauge2}
\end{equation}
There are $n(n+1)/2$-parameters $\varepsilon _{ij}\left( \tau \right) $
which $\,$mix positions with velocities. The inverse $e^{ij}$ transforms as $%
\delta e^{ij}=-e^{ik}\left( \partial _\tau \varepsilon _{kl}\right) e^{lj}.$
The Lagrangian transforms into a total derivative given by 
\begin{equation}
\delta L=\frac 12\partial _\tau \left( \dot{x}_i^\mu \varepsilon ^{ij}\dot{x}
_j^\nu \eta _{\mu \nu }-\left( m^2\right) ^{ij}\varepsilon _{ij}\right) .
\label{dL}
\end{equation}
The action is invariant provided $\varepsilon _{ij}\left( 0\right)
=\varepsilon _{ij}\left( T\right) =0.$ This is a new gauge symmetry that
cannot be thought of as $\tau $ reparametrization\footnote{%
It is possible to combine $\varepsilon _{ij}\left( \tau \right) e^{jk}\left(
\tau \right) $ into a new local parameter $\bar{\varepsilon}_i^k\left( \tau
\right) $. Then for $n=1$ the new transformation becomes the standard $\tau $
reparametrization. However, for $n>1$ it cannot be written as
reparametrizations of a single $\tau $, showing that this is a new gauge
symmetry for multi-particles.\label{repar}} for $n>1$.

A first order formalism may also be given by defining the canonical momenta $%
p_\mu ^i\equiv e^{ij}\dot{x}_j^\mu ,$ and replacing all velocities in the
Hamiltonian by the momenta 
\begin{equation}
S=\int_0^Td\tau \,\left( \dot{x}_i^\mu p_\mu ^i-\frac 12\left[ p_\mu
^i\,p_\nu ^j\,\eta ^{\mu \nu }+\left( m^2\right) ^{ij}\right] e_{ij}\right) .
\end{equation}
This action is invariant under a local multi-particle gauge transformation
given by 
\begin{equation}
\delta x_i^\mu =\varepsilon _{ij}e^{jk}\,\partial _\tau x_k^\mu ,\quad
\delta e_{ij}=\partial _\tau \varepsilon _{ij},\quad \delta p_\mu
^i=e^{ij}\varepsilon _{jk}\,\partial _\tau p_\mu ^k\,,  \label{dp}
\end{equation}
since the variation of the Lagrangian is a total derivative 
\begin{equation}
\delta L=\partial _\tau \left( p_\mu ^i\varepsilon _{ij}e^{jk}\dot{x}_k^\mu
-\frac 12p_\mu ^i\,p_\nu ^j\,\eta ^{\mu \nu }\varepsilon _{ij} -\frac 12
\left( m^2\right) ^{ij}\varepsilon _{ij}\right) .
\end{equation}
We emphasize that, in the first order formalism, the symmetry is valid when
each function $x,e,p$ is varied independently according to (\ref{dp}), {\it %
without} using any equation of motion (such as the relation between momenta
and velocities). There is a true symmetry in the off shell path integral.

According to the equations of motion the canonical momenta are conserved, $%
\partial _\tau p_\mu ^i=0.$ Furthermore, they are constrained as in (\ref
{multi-p}). One can choose a basis in which the mass matrix is diagonal $%
\left( m^2\right) ^{ij}=\delta ^{ij}m_i^2.$ Thus, the new formulation
reproduces the set of constraints that were obtained in the previous
formulations.

The gauge symmetry permits the gauge fixing $e_{ij}=\delta _{ij}$, bringing
the system to the form of a ``gas'' of particles that move freely except for
the fact that their conserved momenta are mutually orthogonal. In this sense
there is an interaction among them.

\section{Superparticles, global and local new SUSY}

For multi-superparticles there are $n$ positions $x_i^\mu \left( \tau
\right) $ and momenta $p^{i\mu }\left( \tau \right) $, $i=1,2,\cdots n$, and 
$N$ spinors $\theta _{\alpha A}\left( \tau \right) $ labelled by $%
A=1,2,\cdots ,N$, where $N$ is the number of supersymmetries, and the index $%
A$ is unrelated to the index $i$. The indices $\mu $,$\alpha $ are
classified as the vector and spinor in $d+2n-2$ dimensions, with spacetime
symmetry SO$\left( d+n-2,n\right) $. We propose an action directly in the
first order formalism\footnote{%
We are discussing dimensions $d+2n-2$ for which $\gamma _{\alpha \beta
}^{\mu _1\cdots \mu _n}$ is symmetric in ($\alpha \beta ).$ For $n=1$ start
with $d=10.$ Then, for $n=2,$ twelve dimensions with signature $\left(
10,2\right) $, and for $n=3,$ fourteen dimensions with signature $(11,3)$
satisfy this property. These are the dimensions of greatest interest in our
program. For lower dimensions see footnote in \cite{spartstr}.} 
\begin{equation}
S=\int_0^Td\tau \,\left( \dot{x}_i^\mu p_\mu ^i-\frac 12p_\mu ^i\,p_\nu
^j\,\eta ^{\mu \nu }\,e_{ij}+\bar{\theta}_A\Gamma \left( p\right) \dot{%
\theta }_A\,\right) .
\end{equation}
where 
\begin{equation}
\Gamma _{\alpha \beta }\left( p\right) =\gamma _{\alpha \beta }^{\mu
_1\cdots \mu _n}\left( p_{\mu _1}^1\cdots p_{\mu _n}^n\right) .  \label{gam}
\end{equation}
The second order formalism, in which the Lagrangian is written only in terms
of velocities is extremely non-linear, and fortunately not needed.

This action is invariant under the following bosonic gauge symmetry that
generalizes the local multi-particle symmetry of (\ref{gauge2}) to
superparticles 
\begin{eqnarray}
\delta x_i^\mu &=&\varepsilon _{ij}e^{jk}\left( \partial _\tau x_k^\mu +\bar{
\theta}_AV_k^\mu \left( p\right) \partial _\tau \theta _A\right) , \\
\delta p_\mu ^i &=&e^{ij}\varepsilon _{jk}\partial _\tau p_\mu ^k,\quad
\delta e_{ij}=\partial _\tau (\varepsilon _{ij}),\quad \delta \theta _A=0, 
\nonumber
\end{eqnarray}
where 
\begin{equation}
V_k^\mu =\frac \partial {\partial p_\mu ^k}\Gamma \left( p\right) =\gamma
^{\mu _1\cdots \mu _n}\,\frac \partial {\partial p_\mu ^k}\left( p_{\mu
_1}^1\cdots p_{\mu _n}^n\right) .  \label{v}
\end{equation}
Note that $\delta \theta _A=0$, unlike the standard $\tau $
reparametrization, illustrating once again that the multi-particle symmetry
is different than $\tau $ reparametrization$^{\ref{repar}}$. As in the
purely bosonic case, there are $\frac 12n(n+1)$ local parameters $%
\varepsilon _{ij}\left( \tau \right) $. Under these variations the
Lagrangian transforms into a total derivative 
\begin{equation}
\delta L=\partial _\tau \left[ \varepsilon _{ij}e^{jk}\left( p^i\cdot \dot{x}
_k+\bar{\theta}_A\left( p^i\cdot V_k\right) \dot{\theta}_A\right) -\frac
12p^i\cdot p^j\varepsilon _{ij}\right] .
\end{equation}
Therefore the action is invariant provided $\varepsilon _{ij}(0)=\varepsilon
_{ij}(T)=0$.

Our action is also invariant under global and local supersymmetries. The
global supersymmetry transformation is of the new type, involving the
product of momenta which appear in $V_i^\mu \left( p\right) $, 
\begin{eqnarray}
\delta _\varepsilon \theta _{A\alpha } &=&\varepsilon _{A\alpha },\quad
\delta _\varepsilon x_i^\mu =-\bar{\varepsilon}_AV_i^\mu \left( p\right)
\theta _A\,\,,\quad  \label{susy} \\
\delta _\varepsilon p_\mu ^i &=&0,\quad \delta _\varepsilon e_{ij}=0, 
\nonumber
\end{eqnarray}
where $\varepsilon _{\alpha A}$ are $\tau $ independent constant spinors.
The $V_i^\mu \left( p\right) $ which appear in the transformation of
particle $i$ is independent of the momentum of the particle $i,$ but depends
on the product of the momenta of all other particles. Thus, when the momenta
of all other particles are frozen, this supersymmetry transformation reduces
effectively to the standard supersymmetry transformation in the single
particle sector. Examples of this mechanism have been discussed in more
detail in earlier papers. Under (\ref{susy}) the Lagrangian transforms into
a total derivative 
\begin{equation}
\delta L=\left( 1-n\right) \,\,\partial _\tau \left( \bar{\varepsilon}
_A\Gamma \left( p\right) \theta _A\right) .
\end{equation}
To show this, we have used $\partial _\tau \Gamma (p)=V_i^\mu \dot{p}_\mu ^i$
and $V_i^\mu p_\mu ^i=n\Gamma (p).$ The closure of the multi-particle
superalgebra is obtained by considering $\left[ \delta _{\varepsilon
_1},\delta _{\varepsilon _2}\right] $ on the various fields, and is given by 
\begin{equation}
\left\{ Q_{A\alpha },Q_{B\beta }\right\} =\left( \gamma ^{\mu _1\cdots \mu
_n}\right) _{\alpha \beta }\,\left( p_{\mu _1}^1\cdots p_{\mu _n}^n\right) ,
\label{qsusy}
\end{equation}
where the right hand side coincides with $\Gamma \left( p\right) $ given in
( \ref{gam}).

Next we consider local $\kappa $ supersymmetry with the transformations 
\begin{eqnarray}
\delta _\kappa x_i^\mu  &=&\delta _\kappa \bar{\theta}_AV_i^\mu \theta
_A\,\,,\quad \delta _\kappa p_\mu ^i=2e^{ij}\bar{\kappa}_A\left( V_{j\mu
}\Gamma \right) \dot{\theta}_A\,\,,  \nonumber \\
\delta _\kappa \bar{\theta}_{A\alpha } &=&-e^{ij}\left[ \dot{x}_{i\mu }+\bar{%
\theta}_AV_{i\mu }\dot{\theta}_A\right] \left( \bar{\kappa}_AV_j^\mu \right)
_\alpha ,\quad   \label{ktransf} \\
\delta _\kappa e_{ij} &=&-(-1)^{n(n-1)/2}4\,Cof\left( M\right) _{ij}\,\bar{%
\kappa}_A\dot{\theta}_A,  \nonumber
\end{eqnarray}
where we have used 
\begin{equation}
\left( \Gamma \left( p\right) \right) ^2=(-1)^{n(n-1)/2}\det (M),
\label{det}
\end{equation}
and $M$ is the $n\times n$ matrix with entries $M^{ij}=(p^i\cdot p^j),$
determinant $\det M$, inverse $M_{ij}$, and cofactor matrix 
\begin{equation}
Cof\left( M\right) _{ij}=\frac 1{\left( n-1\right) !}\varepsilon
_{ii_2\cdots i_n\,}\varepsilon _{jj_2\cdots j_n}\,M^{i_2j_2}\cdots
M^{i_nj_n}=\det M\,\,\,M_{ij}\,\,.  \label{cof}
\end{equation}
With these $\kappa $ supersymmetry transformations the $\kappa $ variation
of the Lagrangian is a total derivative. To verify this invariance we need
the following steps 
\begin{eqnarray}
\delta _\kappa L &=&\partial _\tau \left( \delta _\kappa x_i\right) \cdot
p+\cdots   \nonumber \\
&=&\partial _\tau \left( \delta _\kappa x_i\cdot p\right) -\delta _\kappa
x_i\cdot \partial _\tau p+\cdots  \\
&=&n\,\partial _\tau \left( \delta _\kappa \bar{\theta}_A\Gamma (p)\theta
_A\right) -\delta _\kappa \bar{\theta}_A\left( \partial _\tau \Gamma
(p)\right) \theta _A+\cdots ,  \nonumber
\end{eqnarray}
where we have used the form of $\delta _\kappa x_i$ given in (\ref{ktransf})
and the identities $V_i^\mu p_\mu ^i=n\Gamma (p)$ and $\partial _\tau \Gamma
(p)=V_i^\mu \dot{p}_\mu ^i$. Combining this with the rest of the $\kappa $
variation, and using $\bar{\theta}_A\Gamma \partial _\tau \left( \delta
_\kappa \theta _A\right) =-\partial _\tau \left( \delta _\kappa \bar{\theta}%
_A\right) \Gamma \theta _A$ due to the symmetry of $\Gamma _{\alpha \beta },$
gives 
\begin{eqnarray}
\delta _\kappa L &=&(n-1)\partial _\tau \left( \delta _\kappa \bar{\theta}%
_A\Gamma (p)\theta _A\right) +2\delta _\kappa \bar{\theta}_A\Gamma (p)\dot{%
\theta}_A  \nonumber \\
&&-\frac 12(p_i\cdot p_j)\delta _\kappa e_{ij}+(\dot{x}_{i\mu }+\bar{\theta}%
_AV_{i\mu }\dot{\theta}_A-e_{ij}p_\mu ^j)\,\delta _\kappa p^{i\mu }.
\end{eqnarray}
The first term is just a total derivative, and the other terms cancel each
other after substituting the $\kappa $ transformations (\ref{ktransf}) and
using $\left\{ \Gamma ,V_i^\mu \right\} \dot{p}_\mu ^i=\partial _\tau \left(
\Gamma ^2\right) $ with the relations (\ref{det}), (\ref{cof}). We emphasize
that each field $x,p,\theta ,e$ is independently transformed, without using
any equation of motion that relates them. In particular the equations of
motion that relate the velocities and momenta are not used. So, the
local $\kappa $ invariance is valid in the fully
off-shell path integral, as are all other symmetries discussed in this paper.

This model provides a realization of the generalized supersymmetry (\ref
{qsusy}) in agreement with previous suggestions \cite{stheory} \cite{sezgin1}%
-\cite{periwal}\cite{martinec}\cite{sezgin3}\cite{spartstr}. The current
model improves on \cite{twotimes}\cite{spartstr} by not having global
Lagrange multipliers, and unlike \cite{sezgin3} avoids using
equations of motion in symmetry transformations\footnote{%
In \cite{sezgin3} another formulation involving multiple $\tau _i$
parameters was suggested. But, to have consistent $\tau _i$ dependence in
the transformations laws of various fields, the momenta of the particles had
to be taken as constants (i.e. on shell, $\partial_{\tau _i}p_{\mu}^i=0$).}.

\section{Superparticle constraints and quantization}

The equations of motion for $p_\mu ^i$ are 
\begin{equation}
e_{ij}p_\mu ^j=\dot{x}_i^\mu +\bar{\theta}_A\gamma ^{\mu _1\cdots \mu _n} 
\dot{\theta}_A\,\frac \partial {\partial p_\mu ^i}\left( p_{\mu _1}^1\cdots
p_{\mu _n}^n\right) .  \label{velo}
\end{equation}
This gives a non-linear set of equations from which velocities are
determined in terms of momenta. The equations of motion for $x_i^\mu $
indicate that the canonical conjugate momenta are conserved $\dot{p}_\mu
^i=0.$ Similarly, the canonical conjugate for $\theta _A$ is 
\begin{equation}
\bar{\xi}^A=\bar{\theta}^A\Gamma \left( p\right) ,  \label{xi}
\end{equation}
and the equation of motion for $\theta _A$ indicates that it is time
independent $\partial _\tau \bar{\xi}^A=0.$

A straightforward application of Noether's theorem for the global
supersymmetry (\ref{susy}) shows that $\xi _{A\alpha }$ is the conserved
supercharge 
\begin{equation}
\xi _{A\alpha }=Q_{A\alpha }.
\end{equation}
Let us derive its commutation rules. According to canonical quantization, it
commutes with the momenta $p_\mu ^i$. Furthermore, ignoring constraints, the
naive anticommutation rules are $\left\{ \theta _{A\alpha },\bar{\xi}%
^{B\beta }\right\} =\delta _A^B\delta _\alpha ^\beta {\ .}$ Multiplying from
the left with $\Gamma \left( p\right) $ we obtain 
\begin{equation}
\left\{ \xi _{A\alpha },\bar{\xi}^{B\beta }\right\} =\delta _A^B\,\,\left(
\Gamma \left( p\right) \right) _\alpha ^\beta \,.  \label{xisusy}
\end{equation}
This is identical to the anticommutation rules of the supercharges given in
( \ref{qsusy}). The supercharges are gauge invariant $\delta _\kappa
Q_{A\alpha }=0$ under the on-shell $\kappa $ transformations. This is easily
seen after using (\ref{velo}) in (\ref{ktransf}) and applying the
constraints below. Therefore, the commutation rules (\ref{xisusy}) or (\ref
{qsusy}) are gauge invariant and consistent with the constraints, even
though the naive anticommutation rules for $\left\{ \theta _{A\alpha },\bar{%
\xi}^{B\beta }\right\} $ need modification due to the constraints.

Applying $\gamma \cdot p^j$ on $\xi ^A$ or $Q^A$ and using the momentum
constraints $p^i\cdot \,p^j=0$ we find that there are also fermionic
constraints 
\begin{equation}
\gamma \cdot p^jQ^A=0.
\end{equation}
These constraints are due to local $\kappa $-symmetries and they help remove
fermionic degrees of freedom. As explained below, each superparticle, in the
background of all other superparticles, effectively has the same bosonic and
fermionic degrees of freedom as the standard superparticle in 10 dimensions.

Let us recall the constraints for a simple superparticle in 10 dimensions
with SO$\left( 9,1\right) $ symmetry (specialize our equations to $n=1$ ).
They are 
\begin{equation}
p^2=0,\quad \not{p}Q^A=0,
\end{equation}
where the Majorana-Weyl spinors $\theta _{A\alpha }$ or $Q^{A\alpha }$ have
16 spinor components $\alpha =1,\cdots ,16.$ The fermionic constraint is
solved by 
\begin{equation}
Q^A=\not{p}\theta ^A.
\end{equation}
$Q$ has only 8 unrestricted components because the lightlike $p^\mu $
projects out half of the components of $\theta $. Thus the constraints are
solved by 8 unrestricted transverse bosonic components in $p^\mu $ and 8
unrestricted fermionic components in $Q_\alpha ^A$ or $\theta _\alpha ^A$
(for each $A$). The 8 unrestricted fermions $Q$ satisfy a Clifford algebra
which follows from (\ref{xisusy}) or (\ref{qsusy}). For one supersymmetry
(one $A$), the quantum Hilbert space is labelled by the Clifford states $%
|a,p>$ consisting of $2^4=8_{bosons}+8_{fermions}$. This is the Yang-Mills
supermultiplet in 10 dimensions. It corresponds to a short representation of
the 10-dimensional superalgebra $\left\{ Q_\alpha ,Q_\beta \right\} =\gamma
_{\alpha \beta }^\mu \,\,p_\mu .$

Next consider the $n=2$ case in 12 dimensions, with SO$\left( 10,2\right) $
symmetry. The Majorana-Weyl spinors $Q_\alpha ^A$ have 32 components $\alpha
=1,\cdots ,32$ for each $A,$ while the momenta $p_\mu ^i$ have two timelike
and 10 spacelike components for each $i$. Two timelike dimensions are
necessary to solve the constraints 
\begin{equation}
p_1^2=p_2^2=p_1\cdot p_2=0,\quad \not{p}_1Q^A=\not{p}_2Q^A=0.
\end{equation}
From (\ref{xi}) we may write the solution of the fermionic constraint 
\begin{equation}
Q^A=\not{p}_1\theta _1^A=\not{p}_2\theta _2^A=\not{p}_1\not{p}_2\theta ^A,
\end{equation}
where we have defined $\theta _1^A\equiv \not{p}_2\theta ^A$ and $\theta
_2^A\equiv -\not{p}_1\theta ^A$. Consider the degrees of freedom of particle
1 in the background of particle 2. For a fixed $p_{2\mu }$ that satisfies $%
p_1\cdot p_2=p_2^2=0$, the momentum $p_{1\mu {\ }}$generally has 10
components in an effective SO$\left( 9,1\right) $ subspace orthogonal to the
lightlike direction of $p_{2\mu }$. Furthermore, for the most general $%
\theta ^A$ we see that $\theta _1^A\equiv \not{p}_2\theta ^A$ can have only
16 components that form the spinor representation in the same SO$\left(
9,1\right) $ subspace. The remaining 10 components of $p_1^\mu $ and 16
components of $\theta _1^A$ (for each $A)$ are further restricted by 
\begin{equation}
p_1^2=0,\quad Q^A=\not{p}_1\theta _1^A,\quad \not{p}_1Q^A=0.
\end{equation}
This is the same set of SO$\left( 9,1\right) $ covariant equations satisfied
by the superparticle in 10 dimensions, as above. Hence, we conclude that
superparticle 1, in the background of superparticle 2, behaves just like the
usual 10 dimensional superparticle with a single timelike dimension. Of,
course, the same argument holds for superparticle 2 in the background of
superparticle 1. The treatment is democratic for either particle. This is
the result we wished to have in order to obtain consistency with standard
physics. The cosmological scenario of \cite{twotimes}\cite{spartstr} may
also be applied.

The remaining degrees of freedom in $Q^A$ or $\theta ^A$ are still 8
unrestricted fermions (for each $A$). These satisfy a Clifford algebra that
follows from (\ref{xisusy}) or (\ref{qsusy}). Therefore, the Clifford states
for the full system are $|a,p_1,p_2>,$ which corresponds again to $%
2^4=8_{bosons}+8_{fermions}$ (for one supersymmetry). These can be expressed
covariantly in 12 dimensions, as the states that form {\it short
supermultiplets} of the 12-dimensional superalgebra 
\begin{equation}
\left\{ Q_\alpha ,Q_\beta \right\} =\left( \gamma ^{\mu \nu }\right)
_{\alpha \beta }\,Z_{\mu \nu }+\left( \gamma ^{\mu _1\cdots \mu _6}\right)
_{\alpha \beta }\,Z_{\mu _1\cdots \mu _6}^{+},
\end{equation}
with $Z_{\mu \nu }\,=\frac 12\left( p_{1\mu }\,p_{2\nu }-p_{1\nu }\,p_{2\mu
}\right) $ and $Z_{\mu _1\cdots \mu _6}^{+}=0$. This general form was first
given in \cite{stheory} as a BPS solution of the general 12D superalgebra.
The model presented here provides an explicit realization of this symmetry
in a short representation.

The arguments are similar for higher values of $n.$ The bigger $n$, the
bigger the spinor. For example, for $n=3$ the spacetime group is SO$\left(
11,3\right) $ in 14 dimensions and the Majorana-Weyl spinors $\theta ,Q$
have 64 components \cite{14D}. For three orthogonal lightlike momenta the
solution of the fermionic constraints is $Q=\not{p}_1\not{p}_2\not%
{p}_3\theta $. In the background of particles 2,3 the effective degrees of
freedom of particle 1 are $p_1,\theta _1$ with the constraint $\not{p}_1Q=0$
, where $\theta _1=\not{p}_2\not{p}_3\theta $ and $Q=\not{p}_1\theta _1.$
Since the $\not{p}_2\not{p}_3$ projector reduces the spinor degrees of
freedom by a factor of 1/4, the number of independent components in $\theta
_1$ is 64/4=16. This is the right number for particle 1 to move effectively
in 10 dimensions. The description is fully democratic for every
superparticle. The overall number of unrestricted fermions is again 8. The
Clifford algebra that they form is realized on quantum states $%
|a,p_1,p_2,p_3>$ which describe $2^4=8_{bosons}+8_{fermions}$ (for one
supersymmetry). This is a short multiplet of the SO$\left( 11,3\right) $
covariant superalgebra in 14 dimensions \cite{14D} 
\begin{equation}
\left\{ Q_\alpha ,Q_\beta \right\} =\left( \gamma ^{\mu \nu \lambda }\right)
_{\alpha \beta }\,Z_{\mu \nu \lambda }+\left( \gamma ^{\mu _1\cdots \mu
_7}\right) _{\alpha \beta }\,Z_{\mu _1\cdots \mu _7}^{+},  \label{14susy}
\end{equation}
such that $Z_{\mu \nu \lambda }=p_{[\mu }^1p_\nu ^2p_{\lambda ]}^3$ and $%
Z_{\mu _1\cdots \mu _7}^{+}=0.$ It can be regarded as a BPS state of a
bigger S-theory \cite{14D}.

\section{Remarks}

It is also possible to formulate multi-particle actions with fewer
constraints, such that instead of constraining the momenta of each particle,
only the {\it total} momenta of groups of particles are constrained. In this
sense, the action given in this paper contains $n$ groups, with a single
particle in each group. We will not discuss the generalization in detail to
groups with many particles here, but instead refer the reader to \cite
{spartstr} for a realization of this idea.

Also, a generalization of our current approach to multi-strings or multi
p-branes may be considered. There will be many strings or p-branes, but
there will be only one set of p-volume parameters $\left( \tau ,\vec{\sigma}
\right) $. The string or p-brane version naturally comes with interactions
through the geometry of the p-brane volume. Progress along those lines will
be reported in a future publication.

We think that the ideas in this paper can be connected to Matrix theory \cite
{banks}, by combining them with the outline in \cite{14D} \cite{typeIIB}, to
provide a matrix formulation that is spacetime covariant, duality covariant
and gauge invariant. As a hint, we can show that our superalgebra (\ref
{qsusy}) fits into the framework of Matrix theory. This point can be
illustrated by rewriting (\ref{qsusy}) for $n=3$ in matrix language 
\begin{equation}
\left\{ Q_\alpha ,Q_\beta \right\} =\gamma _{\alpha \beta }^{\mu _1\mu _2\mu
_3}\left[ \frac 1NTr(X_{\mu _1}X_{\mu _2}X_{\mu _3})+\cdots \right]
\label{msusy}
\end{equation}
with the $N\times N$ matrix $X_\mu =J_1p_\mu ^1+J_2p_\mu ^2+J_3p_\mu ^3$.
The $J_i$ correspond to the $N\times N$ representation of the generators of
SU(2) (embedded in the Lie algebra of SU$\left( N\right) $), i.e. $\left[
J_i,J_j\right] =i\varepsilon _{ijk}J_k$. Then (\ref{msusy} ) reduces to (\ref
{qsusy}). This raises the hope that we should be able to express our ideas
in a matrix formulation and also find a more covariant formulation which
explain the successes of Matrix theory so far. A covariant formulation of
matrix theory or S-theory is expected to include the representations of the
generalized superalgebra (\ref{14susy}) of the type described in this paper,
in addition to those that follow from the non-covariant matrix sector
formulated in \cite{banks}.

There should be a multi-local super Yang-Mills type theory that corresponds
to the field theory version of the present first quantized multi-particle
theory. The number of {\it physical} components of the fields is $%
2^4=8_{bosons}+8_{fermions},$ but the fields generally depend on $n$
locations $A_a\left( x_1^{\mu _1},\cdots ,x_n^{\mu _n}\right) $. For $n=1$
the theory can be formulated covariantly as the standard SO$\left(
9,1\right) $ super Yang-Mills theory in 10 dimensions $A_a\rightarrow \left(
A_\mu ,\psi _\alpha \right) $. For 12 or 14 dimensions the covariant theory
has not been constructed so far. However, hints of its existence have been
provided in references \cite{sezgin1}-\cite{periwal}. We interpret the work
of these authors as a non-democratic description of particle 1 in the
background of the other particles, such that the other particles have been
frozen to constant momenta. Because of the freezing of the momenta the
covariance in \cite{sezgin1}-\cite{periwal} is really SO$\left( 9,1\right) $
rather than SO$\left( 10,2\right) $ or SO$\left( 11,3\right) .$ The freezing
corresponds to a Kaluza-Klein type reduction of the covariant multi-local
field theory. What has been missing in the field theory version is the
democratic treatment of all $n$ particles in terms of fields that depend on $%
n$ locations. The field theory must have supersymmetry of the new type as in
eq.(\ref{qsusy}), and must have some generalized gauge invariance that
imitates a Yang-Mills theory. Its generalization to gravity with $%
2^8=128_{bosons}+128_{fermions},$ as suggested in \cite{stheory}, would be
the 12 or 14 dimensional multi-local generalization of supergravity that
would reduce to local 11D supergravity upon the multi-local Kaluza-Klein
reduction. The construction of such a multi-local field theory remains as a
challenge. However, a first attempt with the new supersymmetry and two
times, with partial reduction appropriate to post Big-Bang physics, has been
successful for free fields \cite{superpv}.

\end{document}